# Oxygen reduction reaction electrocatalysts based on FeSn$_{0.5}$ species embedded in hierarchical CN-graphene based supports

Enrico Negro[a,b], Angeloclaudio Nale[a], Keti Vezzù[a], Federico Bertasi[a], Gioele Pagot[a,c], Stefano Polizzi[d], Alberto Ansaldo[e], Mirko Prato[f], Francesco Bonaccorso[e], Iwona A. Rutkowska[g], Pawel J. Kulesza[g], Vito Di Noto[a,b,h,*]

[a] *"Section of Chemistry for Technologies"*, Department of Industrial Engineering, University of Padova, Via Marzolo 1, I-35131 Padova (PD), Italy, in the Department of Chemical Sciences.

[b] Consorzio Interuniversitario Nazionale per la Scienza e la Tecnologia dei Materiali, Italy.

[c] Centro Studi di Economia e Tecnica dell'Energia "Giorgio Levi Cases", Via Marzolo 9, I-35131 Padova (PD), Italy.

[d] Department of Molecular Sciences and Nanosystems and Centre for Electron Microscopy *"G. Stevanato"*, Università Ca' Foscari Venezia, Via Torino 155/B, 30172 Venezia-Mestre, Italy.

[e] Istituto Italiano di Tecnologia, Graphene Labs, Via Morego 30, 16163 Genova, Italy.

[f] Istituto Italiano di Tecnologia, Materials Characterization Facility, Via Morego 30, 16163 Genova, Italy.

[g] Faculty of Chemistry, University of Warsaw, Pasteura 1, PL-02-093 Warsaw, Poland

[h] Material Science and Engineering Department, Universidad Carlos III de Madrid, Escuela Politécnica Superior, Av.de la Universidad, 30, 28911 Leganes, Spain (present address).

* Corresponding Author. E-mail address: vito.dinoto@unipd.it

**Abstract**

This work reports the synthesis, the physicochemical characterization and the electrochemical studies of new electrocatalysts (ECs) for the oxygen reduction reaction (ORR) that: (i) are based on a hierarchical graphene-based support; and (ii) do not comprise platinum. The active sites of the ECs consist of Fe and Sn species stabilized in *"coordination nests"* of a carbon nitride (CN) matrix. The latter exhibits a rough, microporous morphology and acts as a *"shell"* covering a graphene *"core"*. This paper: (i) discusses the role played by Fe as the *"active metal"* in this family of ECs; and (ii) examines in detail how the physicochemical properties and, correspondingly, the electrochemical performance are affected by a suitable activation procedure **A** meant to boost the ORR kinetics. The results lead to an improved fundamental understanding on the features of the active sites, including the impact of both **A** and the pH of the environment in their performance and ORR mechanism. These insights clarify the most desirable features to be included in high-performing ECs belonging to this family, paving the way to the synthesis of next-generation, efficient ECs for the ORR that do not comprise platinum.

# 1. Introduction

As of today, the most fundamental features of the approach followed in the last decades to generate and distribute energy are experiencing a major upheaval [1,2]. The reliance on fossil fuels is to be curtailed, to minimize the emissions of greenhouse gases and thus mitigate global warming; correspondingly, the role of electricity in the generation, distribution and utilization of energy is expected to rise steeply [3]. To achieve these goals the energy infrastructure must be renovated, allowing for a more efficient exploitation of renewable sources [1-3]; the latter objective can be achieved only by developing technologies for the large-scale storage of energy for the power grid [2,4,5]. In all these regards, electrochemical energy conversion and storage (EECS) technologies such as fuel cells and metal air batteries are expected to play a pivotal role [5,6]. In particular, fuel cells (FCs) and metal-air batteries have attracted significant interest owing to their compatibility with the environment and high energy conversion efficiency. In detail, the latter is not limited by the Carnot cycle and can thus be as high as two-three times larger in comparison with competing technologies (*e.g.*, internal combustion engines, ICEs) [7].

The processes exploited in the operation of FCs and metal-air batteries are bottlenecked by the sluggish kinetics of the oxygen reduction reaction (ORR) [8], that introduces large activation overpotentials, $\eta_{ORR}$, of *ca.* 250 mV or more [9]. Accordingly, the development of electrocatalysts (ECs) capable to minimize $\eta_{ORR}$ is a major goal of the research. From a fundamental viewpoint, major efforts are devoted to study the interplay between the features of the active sites for the ORR and its kinetics and mechanism [10-13]. In practice, the ORR ECs affording the best performance are based on platinum-group metals [14], which are present in Earth's crust with a very low abundance and are prone to cause supply bottlenecks [15]. On these bases, the development of highly efficient ORR ECs that do not comprise platinum-group metals has a very high priority worldwide [14,16]. Such systems are customarily indicated as *"Pt-free"* ECs.

The typical *"Pt-free"* EC consists of a carbonaceous matrix, that coordinates species based on Earth-abundant metals (*e.g.*, Fe[13,17], Co[18,19], W[20], Zn[21]) through one or more type of heteroatoms such as B[21], N[17] and S[22], among many others [23]. A multitude of variations exist on this basic theme: in some instances it is claimed that the EC does not include any metal [24]; in others, the EC does not comprise a carbonaceous matrix [25,26]. Unfortunately, in *"Pt-free"* ECs a comprehensive understanding of the details correlating the ORR performance and mechanism with the physicochemical properties of the material is still missing. This hinders the rational development of new, high-performing and durable *"Pt-free"* ECs. On one hand, the identification of the *"real"* ORR active site for every EC is often difficult; on the other hand, the mechanistic details of the ORR are often studied only superficially, with a particular reference to the role played by the pH of the environment. Since: (i) the latter has a major impact on the ORR mechanism [10,11,27]; and (ii) in different EECS devices, the ORR takes place at a different pH (from 0-1 in the case of proton-exchange membrane fuel cells, PEMFCs [28] to 13-14 for anion-exchange membrane fuel cells, AEMFCs [29]), this is a major shortcoming of the state of the art in this field.

The chemical composition and the morphology of the EC support are the most important factors in the modulation of: (i) the electrical conductivity of the material; and (ii) the accessibility of the active sites to reagents. These features play a major role to determine the conversion efficiency of EECS devices [30,31]. This is especially the case as the EECS operates at large current densities (that can be on the order of a few A·cm$^{-2}$ in the case of PEMFCs [14]). Typical *"Pt-free"* ECs adopt mesoporous carbon supports exhibiting a specific area of *ca.* 100-1000 m$^2$·g$^{-1}$ [32,33]. The latter may trigger severe issues ascribed to mass transport; indeed, the pores of such supports are characterized by a very small average size (on the order of a few nm or less), thus the paths that must be followed by reactants and products between the EC surface and the active sites are very tortuous [34]. These issues may be addressed developing ORR ECs comprising a graphene support. Indeed, the latter is characterized by several outstanding properties, *i.e.*: (i) an electron mobility as high as 200000 cm$^2$·V$^-$

$^1\cdot$sec$^{-1}$ [35], corresponding to an exceptional electron conductivity; and (ii) a morphology that, at least in theory, does not exhibit any microporosity and achieves a specific surface area as large as *ca.* 2600 m$^2\cdot$g$^{-1}$ [36] for easy access to the environment. The high chemical inertia of graphene [37] hinders the application of the latter 2D material in the synthesis of ECs. Nonetheless, a number of approaches are adopted to devise graphene-supported *"Pt-free"* ECs, such as: (i) chemical reduction [38,39]; and (ii) hydrothermal processes [40].

The ECs described here, that exhibit a hierarchical *"core-shell"* morphology and comprise a carbon nitride (CN) *"shell"* matrix, include a support *"core"* consisting of graphene and are an early attempt to introduce this 2D material into ECs synthesized with the preparation protocol devised in our laboratory in the last 15 years [41,42]. The latter allows for a good control of the chemical composition and morphology of ECs, that are based on a CN matrix. In particular, the materials described in this report belong to the *"second-generation"* of CN-based ECs [27], which are characterized by a *"core-shell"* morphology. In these systems, a *"core"* consisting of conductive NPs (*e.g.*, carbon black [43-45] or mesoporous carbons [30]) is covered by a CN *"shell"* with a low concentration of N (lower than *ca.* 5 wt%, to minimize the ohmic losses [41]) that embeds/supports the ORR active sites in *"coordination nests"* based on C- and N- ligands [27,30,41]. The outstanding flexibility of the proposed synthetic route enables the preparation of several different types of ECs, comprising either: (i) a low loading of platinum-group metals (L-PGM ECs) [30,44]; or (ii) completely *"Pt-free"* [27].

A cornerstone of the proposed preparation protocol is the synthesis of a precursor that, in *"second-generation"* CN-based ECs, is based on the desired support that goes through impregnation with a zeolitic inorganic-organic polymer electrolyte (Z-IOPE) [46], yielding a *"nanocomposite zeolitic inorganic-organic polymer electrolyte"* (nZ-IOPE). In these systems, a suitable binder (*e.g.*, sucrose) [44,46] crosslinks clusters of the desired metals. Afterwards, the precursor goes through: (i) a multi-step pyrolysis process; and (ii) additional treatments (*e.g.*, washing in H$_2$O, or reaction with HF with a subsequent pyrolysis [27]). All these steps are critical in the modulation of the physicochemical features and electrochemical performance and ORR mechanism of the product; the resulting ECs are capable of an exceptional performance, both in half-cell configuration [45,47] (as determined in studies carried out by cyclic voltammetry with the rotating ring-disk electrode, CV-TF-RRDE) and in single PEMFC running in operating conditions [30,43]. The support *"core"* present in *"second-generation"* CN-based ECs: (i) guarantees that the ohmic drops are minimized, ensuring a facile electron transport between the external circuit and the active sites; and (ii) disperses very effectively the active sites, with the purpose to enhance their accessibility to reactants and products [44,47]. The role of the CN *"shell"* is to stabilize the active sites in *"coordination nests"* based on C- and N-ligands. The specific choice of reactants (with a particular reference to cyanometalates, that are used in the synthesis the Z-IOPE precursor [46,48] and are the only source of N atoms in the system) ensures that the largest fraction of N atoms are found in the *"coordination nests"* of alloy nanoparticles, stabilizing the active sites and, at the same time, not compromising the charge transport properties of the EC [27].

The experience and the results obtained with the ECs described in this report are to be exploited in future experiments, where new ECs will be better able to take advantage of the outstanding physicochemical features of graphene and accomplish an enhanced performance and durability. Two metals are introduced in the ECs described in this work, namely Fe (the *"active metal"*) and Sn (the *"co-catalyst"*) [41]. Fe is selected as the *"active metal"* for its well-known capability to promote the ORR in *"Pt-free"* ECs [13,17,49]. The CN *"shell"* is likely capable to stabilize effectively the Sn *"co-catalyst"*, that is known to form strong and stable bonds with carbon [50]. In the proposed operating conditions in terms of pH, Sn is expected to yield stable oxides [51]; the latter would boost the hydrophilicity of the EC surface. Consequently, the ORR kinetics would be promoted, particularly in an alkaline environment. Indeed, in these latter conditions the first, rate-determining electron transfer of the ORR is most likely mediated by an *"outer-shell"* process between an $O_2$ molecule surrounded by water and the EC surface [12].

## 2. Experimental

*2.1. Reagents*

Potassium hexacyanoferrate(II) trihydrate, reagent grade, is supplied by Sigma-Aldrich; dimethyltin dichloride, 95% is ABCR product. Sucrose, molecular biology grade, is purchased from Alfa Aesar. The pristine graphene nanoplatelets, that will be labeled *"Pristine Gr"* throughout the manuscript, are obtained from ACS Material, LLC. Perchloric acid (67-72%), hydrogen peroxide (35%), and potassium hydroxide (98.4 wt%) are procured from Fluka Analytical, Merck and VWR International, respectively. Nitric acid (>65%), hydrofluoric acid (48 wt%), methyl alcohol (> 99.8 wt%) and isopropanol (> 99.8 wt%) are products of Sigma-Aldrich. In this work, the reference EC is *"EC-10"* of ElectroChem, Inc. The nominal Pt loading of this material is 10 wt%, and is labelled *"Pt/C ref."* throughout the manuscript. Vulcan® XC-72R is provided as a courtesy from Carbocrom s.r.l. This product is treated with $H_2O_2$ (10 vol.%) before use. No further purification procedure is applied to the reagents, that are used as received. Bidistilled water is used in all the experiments.

*2.2. Synthesis of the Gr-supported ECs*

*2.2.1. Synthesis of Gr support*

2 g of Pristine Gr are suspended into 30 mL of $HNO_3$ under vigorous stirring. 30 mL of $H_2O_2$ are added dropwise. The obtained suspension is stirred 12 h at room temperature, then filtered on a Büchner funnel and extensively washed until the pH of the mother waters reaches 7. The resulting solid is dried in a ventilated oven at a temperature of 150°C. This process yields the graphene nanoplatelet support, henceforth labelled *"Gr"*.

*2.2.2. Synthesis of FeSn$_{0.5}$ CN Gr-supported ECs*

Sucrose (300 mg) is transferred into a beaker made of Teflon® and dissolved into *ca.* 40 mL of methanol; Gr (300 mg) is added to the resulting solution, yielding a dispersion that is homogenized with a probe sonicator (Sonoplus Bandelin HD 2200; duty cycle 0.2, 2 minutes). The dispersion is stirred and heated to *ca.* 60°C by means of an oil bath; this system is brought to *ca.* 2 mL volume, then it is allowed to reach room temperature. Approx. 0.5 mL of water are used to dissolve 125 mg of potassium hexacyanoferrate(II) trihydrate; the resulting solution is added dropwise to the dispersion, then the product is homogenized using the probe sonicator. Approx. 0.5 mL of water are used to dissolve 34 mg of dimethyltin dichloride; the resulting solution is added dropwise to the above dispersion. The obtained dispersion, that goes through extensive homogenization with the probe sonicator, is then stirred for 24 h and then allowed to rest for 24 h rest; the product is subsequently dried in a ventilated oven at 120°C. A solid pellet of precursor is obtained. The pellet then goes through a pyrolysis process under dynamic vacuum of *ca.* 1 mbar, as follows: (i): 150°C, 7 hours; (ii): 300°C, 2 hours; (iii) 900°C, 2 hours. The product is divided into two parts. One part is treated three times with water and dried in a ventilated oven at 120°C, yielding the *"pristine EC"* labeled *"FeSn$_{0.5}$-CN$_l$ 900/Gr"* in accordance with the nomenclature proposed elsewhere [41]. One fraction of pristine EC goes through an etching treatment with 10 wt% HF lasting 2 h, then it is thoroughly washed with water. The powders then goes through a 2 h pyrolysis process at 900°C (in a dynamic vacuum of *ca.* 1 mbar). The product of this activation process (**A**) is the *"activated EC"* that is labeled *"FeSn$_{0.5}$-CN$_l$ 900/Gr$_A$"*.

## 3. Results and Discussion

*3.1. Synthesis Observations*

The synthetic route adopted to obtain the ECs described in this report includes the following three steps: (i) a precursor is prepared; (ii) the precursor undergoes a multi-step pyrolysis process; (iii) the resulting product goes through additional treatments and, optionally, an activation process **A** meant to boost the performance in the ORR [41].

In this report, the precursor consists of a *"nanocomposite zeolitic inorganic-organic polymer electrolyte"* (nZ-IOPE). Similarly to a conventional zeolitic inorganic-organic polymer electrolyte (Z-IOPE) [46], the nZ-IOPE includes: (i) anionic complexes (obtained from the reaction of $Sn(CH_3)_2Cl_2$ and $Fe(CN)_6^{4-}$ present in the starting reaction mixture) that are bridged by (ii) organic species (*i.e.*, sucrose and the Gr support). nZ-IOPEs are obtained through a series of well-assessed equilibria [44,48], that involve the substitution of the chloride ligands on the anionic complexes with the hydroxyl groups found on either the sucrose or the Gr support [44]. The presence of hydroxyl groups on the surface of the Gr support is witnessed by the high-res XPS profiles (see Figure 6). The heteroatoms of the nZ-IOPE (*i.e.*, H, O and N) are for the most part expelled during the multi-step pyrolysis process [44]. The latter yields a system characterized by a *"core-shell"* morphology; the Gr support acts as a *"core"*, that is covered by a well-graphitized carbon nitride (CN) *"shell"*[27]. During the pyrolysis process, metal-based species are also nucleated into the CN *"shell"* [44]. One important feature of the proposed synthetic route is that in the starting nZ-IOPE the cyano ligands (that are the only source of N) develop covalent bonds with the metal species [44,46,48]. Correspondingly, after the pyrolysis process the cyano groups merge into the CN *"shell"* [52,53], and stabilize the metal species of the EC in C- and N-based *"coordination nests"*. The metal species stabilized into the *"coordination nests"* probably play a crucial role to determine the performance of the EC in the ORR [27]. Indeed, the post-pyrolysis treatment and **A** are meant: (i) to remove contaminants and *"inert"* species; and (ii) facilitate the transport of reactants and products, to ensure that such stabilized metal species are able to operate as effectively as possible, maximizing the ORR performance.

*3.2. Chemical composition*

The chemical composition of the ECs in the bulk and on the surface is evaluated respectively by: (i) inductively-coupled plasma atomic emission spectroscopy (ICP-AES) and microanalysis, see Table 1; and (ii) x-ray photoelectron spectroscopy (XPS), see Table 2. The molar ratios $n_H/n_C$, $n_O/n_C$, $n_N/n_C$, $n_{Fe}/n_C$, $n_{Fe}/n_N$, $n_{Sn}/n_{Fe}$, of the ECs are shown in Figure 1.

Table 1. Bulk chemical composition of the $FeSn_{0.5}$ Gr-supported ECs and the supports.

| Electrocatalyst | $K^{(a)}$ | $Fe^{(a)}$ | $Sn^{(a)}$ | $C^{(b)}$ | $H^{(b)}$ | $N^{(b)}$ | Formula |
|---|---|---|---|---|---|---|---|
| $FeSn_{0.5}$-$CN_l$ 900/Gr | 0.05 | 3.44 | 1.55 | 86.0 | 0.36 | 0.55 | $K_{0.02}[FeSn_{0.21}C_{116.3}H_{5.80}N_{0.64}]$ |
| $FeSn_{0.5}$-$CN_l$ 900/$Gr_A$ | 0.02 | 0.38 | 0.47 | 91.7 | 0.09 | 0.28 | $K_{0.06}[FeSn_{0.59}C_{1125}H_{13.2}N_{2.95}]$ |
| Gr | - | - | - | 98.2 | - | - | C |
| Pristine Gr | - | - | - | 99.6 | 0.3 | - | $C_{285}H_{10.2}$ |

(a) Determined by ICP-AES
(b) Determined by microanalysis

Table 2. Surface chemical composition of the $FeSn_{0.5}$ Gr-supported ECs and the supports.

| Electrocatalyst | Fe | Sn | C | O | N | Formula |
|---|---|---|---|---|---|---|
| $FeSn_{0.5}$-$CN_l$ 900/Gr | 0.29 | 0.09 | 96.6 | 2.3 | 0.7 | $FeSn_{0.31}C_{333}O_{7.9}N_{2.4}$ |
| $FeSn_{0.5}$-$CN_l$ 900/$Gr_A$ | $-^{(a)}$ | $-^{(a)}$ | 99.05 | 0.95 | $-^{(a)}$ | $C_{333}O_{3.2}$ |
| Gr | - | - | 97.0 | 3.0 | - | $C_{880}O_{27.2}$ |
| Pristine Gr | - | - | 96.5 | 3.5 | - | $C_{880}O_{31.9}$ |

(a) This value is lower than the detection limit of the XPS instrumentation.

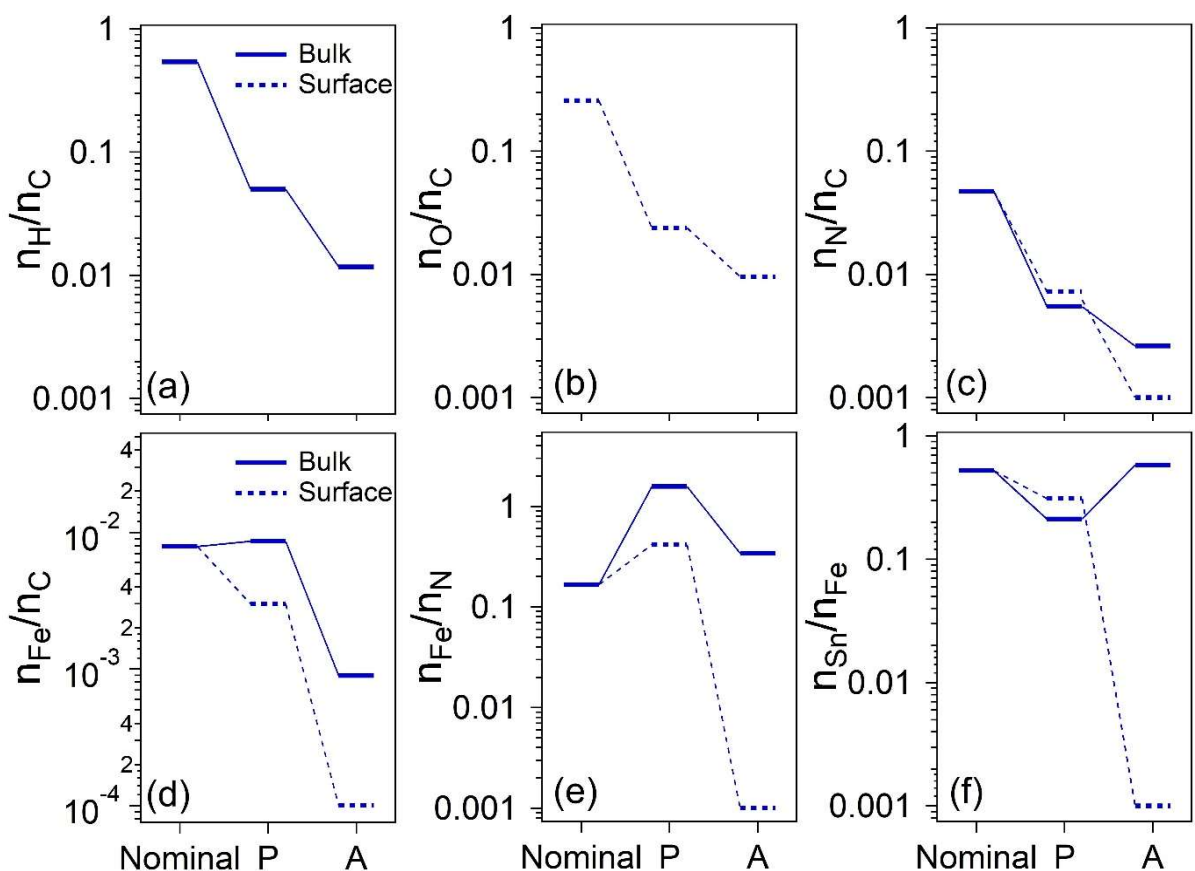

**Figure 1.** Chemical composition of the FeSn$_{0.5}$ Gr-supported ECs. The figure reports the following molar ratios: $n_H/n_C$ (a); $n_O/n_C$ (b); $n_N/n_C$ (c); $n_{Fe}/n_C$ (d); $n_{Fe}/n_N$ (e); $n_{Sn}/n_{Fe}$ (f). The labels at the bottom of each panel are: *"Nominal"* (evaluated on the stoichiometry of the reagents); *"P"* (pristine EC, FeSn$_{0.5}$-CN 900/Gr); and *"A"* (activated EC, FeSn$_{0.5}$-CN 900/Gr$_A$).

The overall trends exhibited by the $n_H/n_C$, $n_O/n_C$ and $n_N/n_C$ molar ratios (see the panels (a), (b) and (c) of Figure 1) are remarkably similar to one another, as follows: (i) with respect to the nominal values, in pristine ECs the ratios are lower by *ca.* 1 order of magnitude; and (ii) the ratios further decrease by a factor of 4-8 upon **A**. These results reveal that the multi-step pyrolysis process and **A** are very effective at removing from the system the heteroatoms (*i.e.*, H, O, and N), yielding a well-graphitized CN *"shell"*. The concentration of N in FeSn$_{0.5}$-CN$_l$ 900/Gr$_A$: (i) in the bulk, it is equal to 0.28 wt% (see Table 1); and (ii) on the surface, it is negligible (see Table 2). Accordingly, **A** etches practically all the N on the external EC surface (*i.e.*, the only part of the sample that can be inspected by XPS [54]); at the same time, N is still preserved elsewhere in the CN *"shell"*, probably both in its bulk and on the surface of its pores (see Section 3.4).

The bulk $n_{Fe}/n_C$ ratio: (i) for the pristine EC, it is quite close to the nominal values; and (ii) it decreases by *ca.* 1 order of magnitude upon **A** (see Figure 1(d)). These results indicate that the proposed synthetic protocol allows for the preparation of ECs with a well-controlled content of metals [27,48], that are mostly etched by **A** [27]. The bulk $n_{Fe}/n_N$ ratio: (i) for the pristine EC, it is *ca.* 1 order of magnitude larger than the nominal values; and (ii) it decreases by a factor of *ca.* 8 upon **A** (see Figure 1(e)). This evidence is consistent with: (i) a substantial, selective removal of N in comparison to Fe during the multi-step pyrolysis process; (ii) a low degree of interaction of Fe by the N atoms of the *"coordination nests"* in the pristine EC; (iii) a selective etching of Fe in comparison to N during **A**; however, in the activated EC, a larger fraction of the surviving Fe species are stabilized by N atoms of the *"coordination nests"*. The bulk $n_{Sn}/n_{Fe}$ ratio: (i) in the pristine EC, it is *ca.* 1/3 of the nominal value; and (ii) upon **A**, it rises by a factor of *ca.* 3 (see Figure 1(f)). These figures suggest that Sn: (i) is removed during the various steps of the preparation route yielding the pristine EC (*e.g.*, pyrolysis and treatment in $H_2O$, see Section 2.2); and (ii) with respect to Fe, it is less easily removed during **A**, probably owing to the formation of stable Sn-C bonds with the CN *"shell"* [50,55]. The $n_{Fe}/n_C$, $n_{Fe}/n_N$ and $n_{Sn}/n_{Fe}$ ratios on the surface of the ECs are typically lower than in the bulk (see the panels (d), (e) and (f) of Figure 1). This result is attributed to: (i) the presence of carbonaceous species that partially cover Fe and lower its surface concentration in the pristine EC, similarly to what is reported in the literature [27]; and (ii) upon **A**, a practically complete etching of Fe and Sn species from the surface of the activated EC.

*3.3. High-resolution thermogravimetric studies*

The results of the investigations carried out by high-resolution thermogravimetry (HR-TGA) are displayed in Figure 2. The features of the thermal events in an oxidizing atmosphere are reported in Table 3.

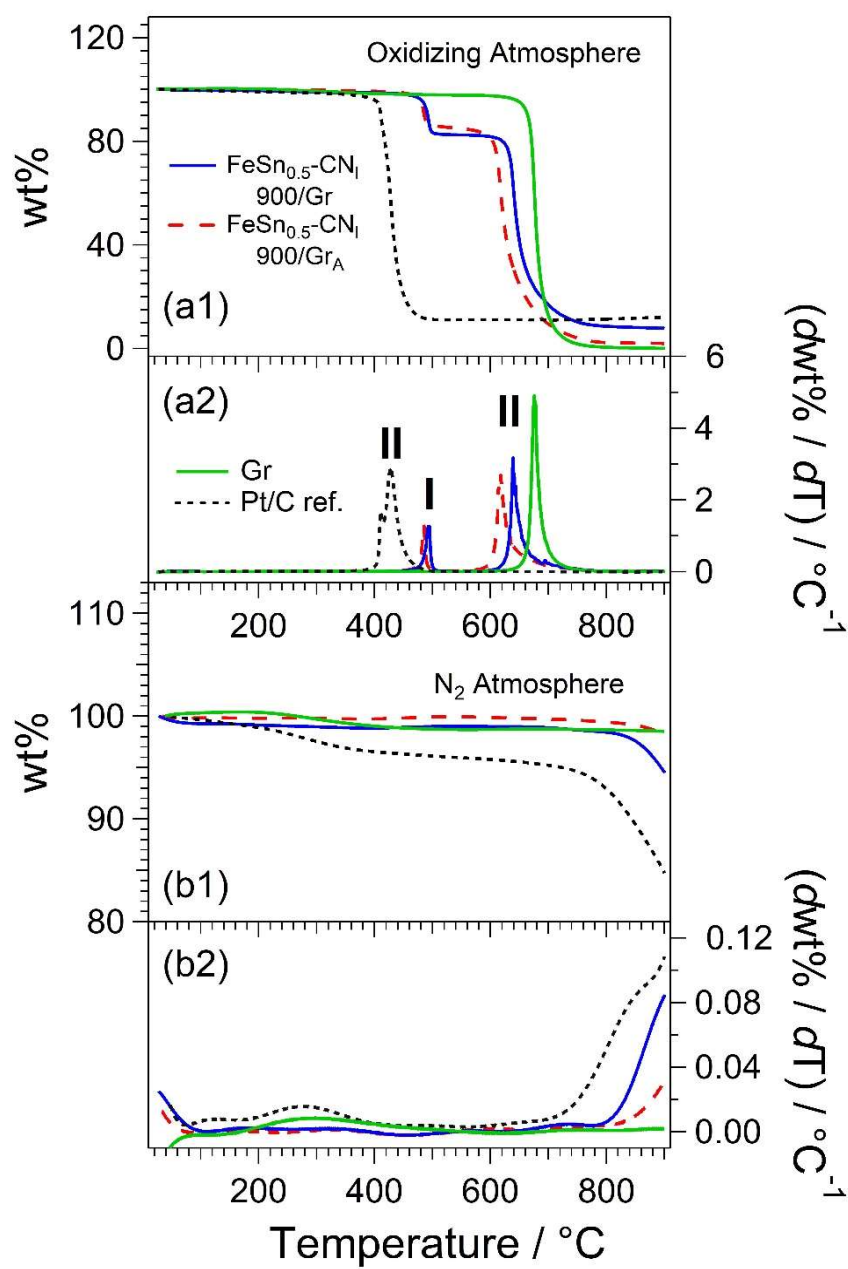

**Figure 2.** HR-TGA profiles of the FeSn$_{0.5}$ Gr-supported ECs in both an oxidizing atmosphere (a1) and in a N$_2$ atmosphere (b1). The corresponding derivatives are plotted in (a2) and (b2), respectively.

**Table 3.** Thermal features of the HR-TGA profiles in an oxidizing atmosphere displayed in Figure 2.

| Sample | Event I | | Event II | | Residual / Wt% |
|---|---|---|---|---|---|
| | T / °C | ΔWt / % | T / °C | ΔWt / % | |
| FeSn$_{0.5}$-CN$_l$ 900/Gr | 494 | 17.4 | 639 | 74.8 | 7.8 |
| FeSn$_{0.5}$-CN$_l$ 900/Gr$_A$ | 484 | 14.7 | 617 | 83.2 | 2.1 |
| Pt/C ref. | | | 427 | 88.8 | 11.2 |
| Gr | | | 676 | 99.5 | 0.1 |

Two main events are revealed in the HR-TGA profiles under an air oxidizing atmosphere, indicated as I and II. I corresponds to a mass loss between *ca.* 15 and 18 wt%; it occurs in the range 460 < $T_I$ < 510°C, and is assigned to the oxidative degradation of the CN *"shell"*. II is associated to a mass loss between *ca.* 75 and 85 wt%; it takes place at 600 < $T_{II}$ < 650°C, and is assigned to the combustion of the Gr support. This latter is validated considering that the Gr support is subjected to degradation at $T_{II}$ = 676°C (see Table 3). The corresponding event revealed in the Pt/C ref. (*i.e.*, oxidative decomposition of its XC-72R support) is shifted to a lower temperature, $T_{II}$ = 427°C (see Table 3), owing to the catalytic effect of the Pt nanoparticles [56]. The residue detected at high temperatures for the proposed ECs is equal to *ca.* 8 and 2 wt% for the pristine and the activated ECs, respectively. The residue is ascribed to stable oxide/carbon/nitride species including Fe and Sn remaining after the oxidative degradation of the carbonaceous species of both the Gr support *"core"* and of the CN *"shell"* [27]. This interpretation is further supported considering that the magnitude of the high-T residue determined by HR-TGA is comparable with the sum of the Fe and Sn wt% detected by ICP-AES (see Table 1).

The following trends are revealed in the HR-TGA profiles under an oxidizing atmosphere (see Figure 2(a1) and Figure 2(a2)): (i) $T_I$ is hardly affected by **A**; and (ii) $T_{II}$ decreases in the order: Gr support > pristine EC > activated EC. The low effect of **A** on $T_I$ is interpreted admitting that: (i) the oxidative decomposition of the CN *"shell"* is triggered by the adsorption of the oxygen included in the air; thus (ii) both the pristine and the activated EC are expected to share very similar sites where oxygen can undergo adsorption on the CN *"shell"*. Accordingly: (i) the Fe and Sn species etched upon **A** do not strongly affect the interactions between oxygen and the CN *"shell"*, thus, they are mostly inert; and (ii) the adsorption of oxygen on the CN *"shell"* is modulated by species that survive after **A**, likely Sn- and Fe-based systems stabilized in the C- and N- ligands of the *"coordination nests"*. The Fe/Sn-based residues left after the combustion of the CN *"shell"* of both the pristine and the activated EC probably facilitate the adsorption of oxygen, lowering $T_{II}$ in comparison with the pristine Gr support (see Table 3). The lowering in $T_{II}$ is more pronounced in the activated EC, where *"almost inert"* Fe/Sn-base species have been removed and no longer inhibit the operation of more active systems (*i.e.*, likely those resulting from the degradation of the Sn- and Fe-based species originally stabilized in the C- and N- ligands of the *"coordination nests"*). The inspection of the HR-TGA traces collected under an inert atmosphere (see Figure 2(b1)) reveals that the proposed ECs: (i) adsorb a very low amount of atmospheric moisture, on the order of 1 wt% of less; and (ii) they are stable up to at least 800°C. The lowest moisture adsorption and the highest thermal stability are detected for the activated EC. This outcome is justified if we consider that the activated EC includes the best-graphitized CN *"shell"*, comprising the lowest amounts of: (i) oxygen; (ii) other heteroatoms; and (iii) metal-based species (see Table 1 and Section 3.2). Indeed, all the latter: (i) increase the hydrophilicity of the EC, facilitating the adsorption of moisture; and (ii) act as defects, promoting the high-temperature oxidative degradation triggered by the oxygen species included in the ECs.

*3.4. Porosimetry and surface structure studies*

Figure 3 reports the specific areas of the pore structural features.

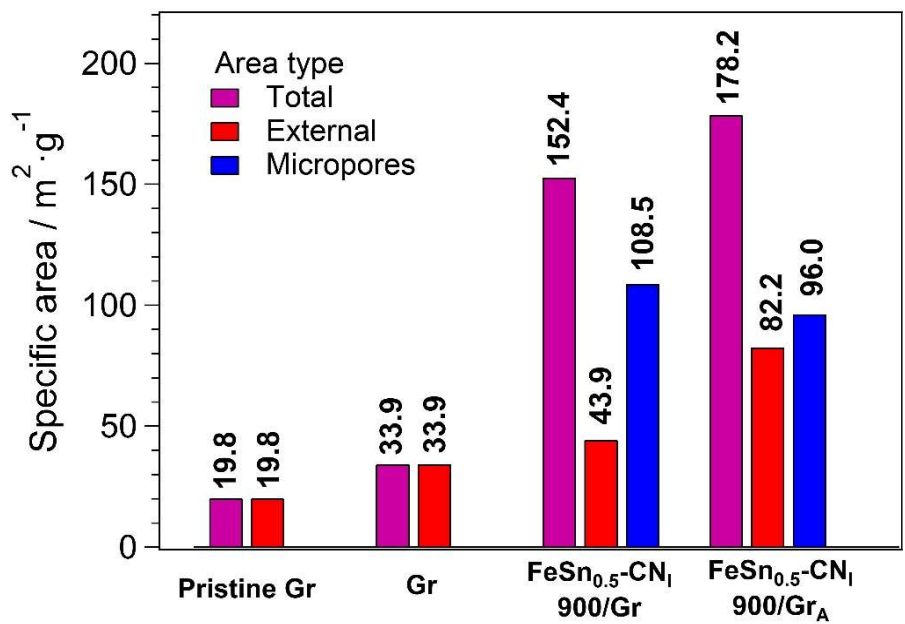

**Figure 3.** Surface area of the hierarchical FeSn$_{0.5}$ Gr-supported ECs as determined by nitrogen physisorption techniques.

The supports *"Pristine Gr"* and *"Gr"* do not exhibit any microporosity, consistently with the sheet-like 2D structure of graphene [36]. The total specific area of Pristine Gr and of the Gr supports is equal to 19.8 and 33.9 m$^2$·g$^{-1}$, respectively. These results suggest that: (i) the proposed synthetic route (see Section 2.2) is able to partially exfoliate Pristine Gr; and (ii) nevertheless, the Gr support consists of stacked graphene layers. Indeed, even if the *"ideal"* graphene monolayer would exhibit a specific area of *ca.* 2600 m$^2$·g$^{-1}$ [36], in practical systems the specific area of graphene-like sheets can reach 700-800 m$^2$·g$^{-1}$ [57,58]. With respect to the Gr support, the total specific area of the ECs is much larger, on the order of *ca.* 150-180 m$^2$·g$^{-1}$. In addition, the ECs are microporous. It is thus inferred that, as the ECs are prepared, a combination of the following phenomena takes place: (i) the Gr support *"core"* is further exfoliated; and (ii) the Gr support *"core"* is covered with a rough, microporous CN *"shell"*. The impact of **A** on the pore structural features is the following: (i) the external area almost doubles, increasing from 43.9 to 82.2 m$^2$·g$^{-1}$; and (ii) the micropore area decreases from 108.5 to 96.0 m$^2$·g$^{-1}$. These outcomes are rationalized assuming that the etching of *"inert"* Sn- and Fe-based species from the CN *"shell"* (see Section 3.2 and Section 3.3) leaves behind large, rough craters. The latter can easily be reached by the *"probe"* N$_2$ molecules during the physisorption experiments, and thus contribute to the *"external"* area of the activated EC. It can also be envisaged that as the *"inert"* Sn- and Fe-based species are etched, they bring with them a small fraction of the CN *"shell"*, thus destroying a small fraction of the micropores.

The above picture is corroborated considering the confocal μ-Raman spectra of the supports and ECs (see Figure S1). In particular, the *"$I_D/I_G$"* ratio between the intensities of the D- and the G-bands (that are detected at *ca.* 1340 and 1580 cm$^{-1}$, respectively) is a good figure of merit to gauge the degree of disorder on a carbonaceous system [59,60]. Indeed, the $I_D/I_G$ ratio rises as the density of defects is increased [59,60]. Both supports (*i.e.*, Pristine Gr and Gr) exhibit $I_D/I_G$ ratios that are very low, *ca.* 0.04-0.05 (see Table S1). Accordingly, it is concluded that Pristine Gr and Gr consist of graphene sheets characterized by a low defect density [59,60]. The $I_D/I_G$ ratio of the pristine and the activated EC is equal to 0.233 and 0.273, respectively (see Table S1). This evidence suggests that the degree of disorder of the EC surface: (i) is much higher in comparison with both supports; and (ii) is further raised by **A**. These results are typical of systems covered by a rough, porous CN matrix [52]; in this case: (i) the introduction of N atoms into the system gives rise to a breakdown of the k-selection rule, raising the intensity of the D-band [61]; and (ii) further defects are introduced owing to the roughness of the material. The latter is expected to raise upon **A**, owing to the formation of craters on the CN *"shell"* (see above); as a result, the $I_D/I_G$ ratio of the activated EC is the largest.

## 3.5 Morphology and structural investigations

The morphologies of both the pristine and the activated ECs are depicted in Figure 4; the latter also includes the electron diffraction (ED) patterns of selected representative portions of the ECs.

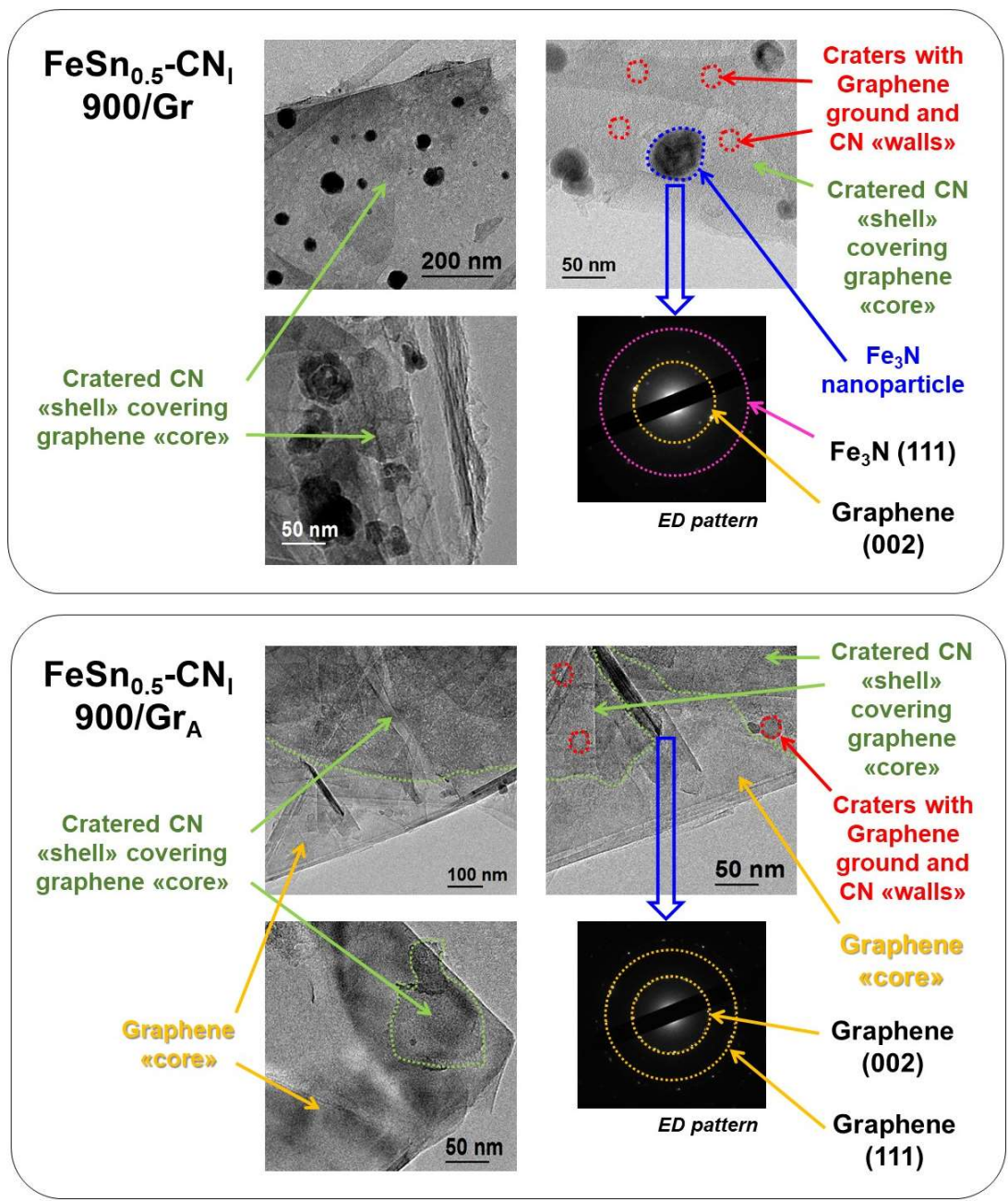

**Figure 4.** HR-TEM micrographs of Gr-supported ECs. Selected-area electron diffraction (ED) patterns provide additional structural information.

The following morphology features are detected for the pristine EC: (i) Dark domains are found (d ≈ 30-50 nm), which are attributed to metal-rich grains based on Fe and/or Sn; (ii) a light matrix is revealed, which is assigned to the Gr support *"core"* covered with the rough CN *"shell"*; and (iii) the CN *"shell"* is not continuous; indeed, it is wrinkled, porous, and dotted with caters. The bottom of such craters probably reach a graphene layer of the Gr support *"core"*; the walls of the craters consist of edges of the CN *"shell"*. The formation of the craters likely takes place during the pyrolysis process, as some parts of the CN *"shell"* are removed as flakes. The ED patterns are consistent with the following two phases: (i) stacked graphene sheets (S.G. *P*6$_3$/mc, COD#9008569) [62]; and (ii) Fe$_3$N (S.G. P312, COD#4000973 [62]; see below for additional details). The morphology of the proposed ECs is strongly affected by **A**. The most evident effect is the complete removal of the dark features. The activated EC exhibits a very rough morphology, where some additional part of the CN *"shell"* matrix has been removed. The surviving portions of the CN *"shell"* are still heavily cratered and rough, and exhibit a larger part of the Gr support *"core"* underneath. However, no additional tiny cavities are formed upon **A**. These results are in accordance with the pore structural features discussed in Section 3.4. The ED patterns of the activated EC further support this picture, revealing features that can be qualitatively ascribed to the same stacked graphene sheets detected in the pristine EC (S.G. *P*6$_3$/mc, COD#9008569) [62]. Thus, we can conclude that **A** does not strongly affect the stacking of the graphene sheets in the Gr support *"core"*.

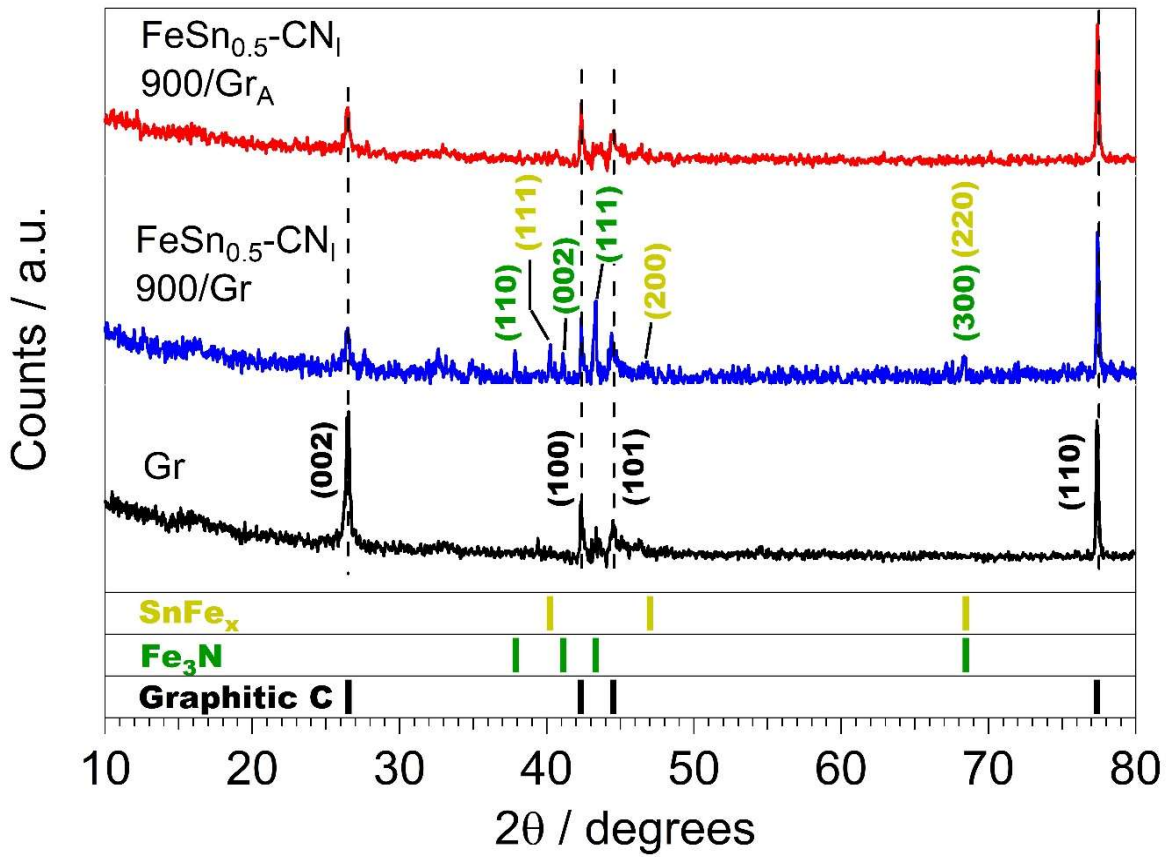

**Figure 5.** Wide-angle X-ray diffraction (WAXD) patterns of the Gr-supported ECs and of the Gr support.

The WAXD pattern of the Gr support exhibits four main reflections at 2θ ≈ 26.5, 42, 44.5 and 77.5°, corresponding to the (002), (100), (101) and (110) peaks of graphitic C (S.G. $P6_3/mc$, COD#9008569) [62]. The Scherrer formula [63] is applied to the (002) reflection; accordingly, the domain size along the [001] direction is evaluated at *ca.* 25 Å. The same procedure is adopted to study the (110) reflection; in this latter case, it is found that the domain size in the planes normal to the [001] direction is larger than 90 Å. These results reveal that the Gr support is strongly anisotropic, including broad graphene sheets (d > 90 Å) that form stacks with an average thickness of 7-8 layers. This latter value is determined considering that the average distance between two graphene sheets in graphite is *ca.* 3.35 Å (see COD#9008569 [62]). It is pointed out that in *"ideal"* crystalline graphite, the ratio between the intensities of the (002) and (110) peaks (hereafter, this figure is indicated as *"$I_{002}/I_{110}$* *ratio"*) is on the order of *ca.* 30 [64]. In the case of Gr support, the $I_{002}/I_{110}$ ratio is much lower, *ca.* 1; this further confirms the stacked sheet structure of this sample. The reflections detected for the Gr support are also revealed in the WAXD patterns of both the pristine and the activated ECs. Furthermore: (i) the full width at half maximum (FWHM) of each (002) and (110) peak is also practically the same for the three materials; and (ii) in both the pristine and the activated EC the $I_{002}/I_{110}$ ratios are lower than in the case of Gr support, on the order of *ca.* 0.25 – 0.35. This outcome is interpreted admitting that the proposed synthetic route: (i) does not affect significantly the long-range ordering of the single graphene layer provided by the Gr support and included in the final ECs; (ii) likely promotes the exfoliation of the stacked graphene sheets included in the Gr support; and (iii) reduces the intensity of the (002) reflection of the Gr support *"core"*. Indeed, the exposed (110) planes of the latter are covered with the CN *"shell"*, that is expected to absorb or scatter a fraction of the X-ray radiation.

The WAXD pattern of the pristine EC also reveals reflections that are assigned as follows: (i) signals at $2\theta \approx 38, 41.5, 43.5$ and $68.5°$, assigned to the (110), (002), (111) and (300) peaks of a $Fe_3N$ phase, S.G. P312, COD#4000973 [62], domain size *ca.* 40 nm; and (ii) signals at $2\theta \approx 40, 47$ and $68.5°$, assigned to the (111), (200) and (220) peaks of a $SnFe_x$ phase, S.G. Fm-3m, COD#9008469 [62], domain size *ca.* 50 nm. The latter phase is a random alloy of cubic Fe and Sn; by applying the mixing law between cubic Fe (COD#9008469) with an α-Sn phase (COD#9008568) [62], it is possible to evaluate x as *ca.* 9. The domain size of $Fe_3N$ and $SnFe_9$ is determined with the Scherrer formula [63]. No additional phases are clearly detected; in particular, no peak ascribable to $FeO_x$ or $SnO_x$ species is revealed. During the preparation process the Fe atoms are introduced as $Fe(CN)_6^{4-}$ complexes (see Section 2.2), that are also the only source of N atoms in the proposed ECs. Accordingly, Fe and N atoms are found in close proximity during the pyrolysis process that triggers the nucleation and growth of the metal-rich domains, and can easily combine forming the $Fe_3N$ phase. Metal nitride phases are also detected in other ECs obtained by our group with a similar preparation route [44].

The $Fe_3N$ and $SnFe_9$ phases observed in the pristine EC are consistent with the dark features evidenced by HR-TEM (see Figure 4), both in terms of domain size and of structure (as witnessed by ED, see Figure 4). It is further pointed out that the $Fe_3N$ and $SnFe_9$ phases are not present in the WAXD pattern of the activated EC. This evidence is consistent with: (i) the chemical composition of the activated EC (see Table 1), that exhibits a dramatic drop in the wt% of Sn and Fe upon **A**; and (ii) the disappearance of dark features in the HR-TEM micrographs of the activated EC (see Figure 4).

*3.6 High-res XPS studies*

The oxidation states of the elements on the surface of the Gr support and of the proposed ECs are studied by high-res XPS. The decomposition of the high-res XPS N 1s and O 1s profiles is displayed in Figure 6. All the spectra are normalized on the integrated area of the C 1s profile, since it is the strongest peak and its surface concentration is practically constant in all the samples (see Table 2). Additional information on the decompositions (*i.e.*, position and relative intensity of the different components) is provided in Table S2 (N 1s profiles) and Table S3 (O 1s profiles).

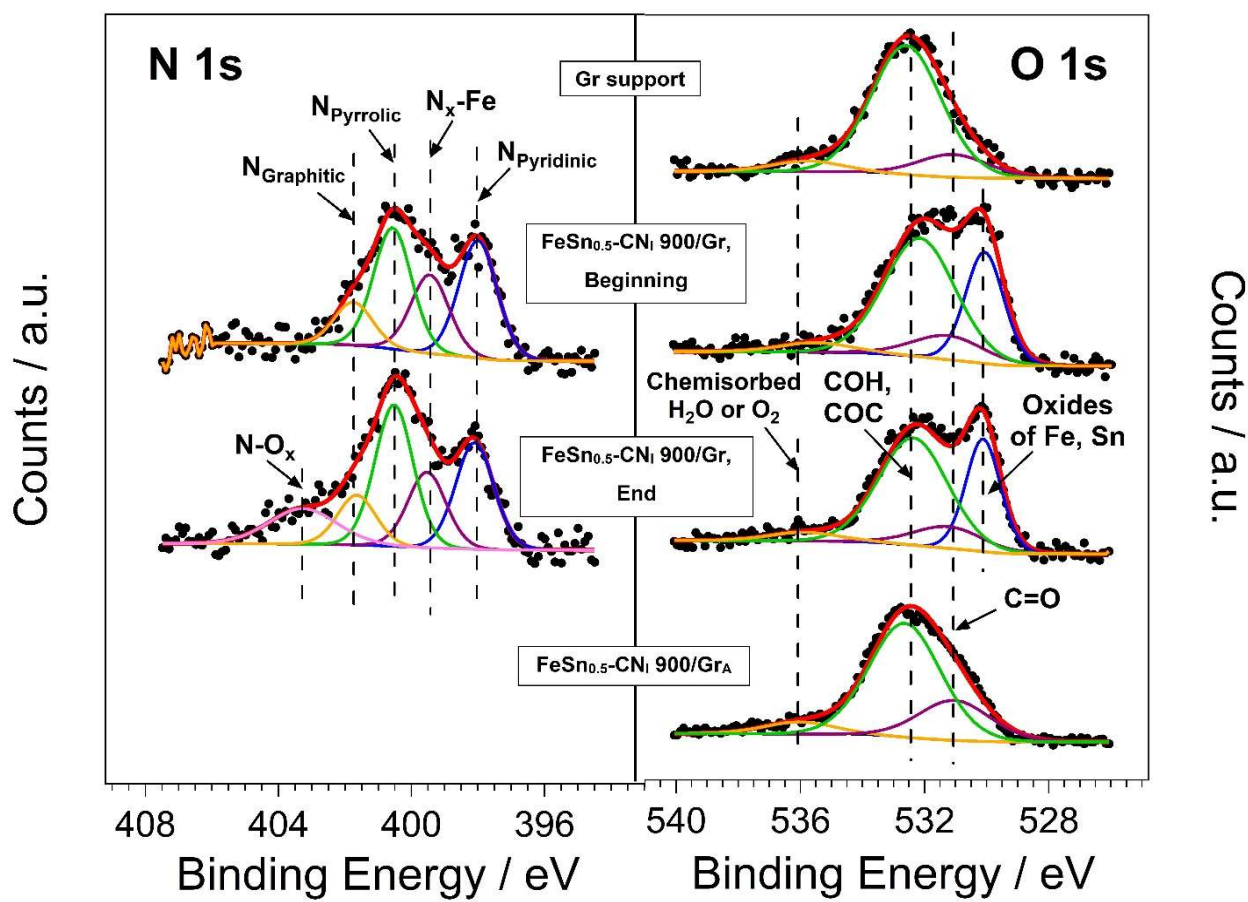

**Figure 6.** Decomposition of the high-res XPS profiles: N 1s (left) and O 1s (right).

The profiles reported in Figure 6 and indicated with the labels *"FeSn$_{0.5}$-CN$_l$ 900/Gr, Beginning"* and *"FeSn$_{0.5}$-CN$_l$ 900/Gr, End"* are collected at the beginning and at the end of the XPS measurements, respectively (the experimental details are reported in Electronic Supplementary Information, ESI). Thus, it is revealed that the high-res N 1s and O 1s profiles evolve during the XPS measurements. No such evolution is detected in the high-res XPS profiles of the other elements (*i.e.*, C, Fe and Sn), that are stable over the course of the measurements (see Figure S3, Figure S4, Figure S5). In detail: (i) the high-res XPS profiles of carbon exhibit almost exclusively a very strong component ascribed to graphitic C (see Figure S3), while the high binding energy *"tail"* ascribed to oxidized species is much less pronounced; and (ii) the high-res XPS profiles of Fe and Sn are compatible with FeO$_x$ ($1 \leq x \leq 1.5$, see Figure S4) and SnO$_x$ ($2 \leq x \leq 4$, see Figure S5) species. It is further pointed out that such oxide species are expected to cover as thin layers the Fe$_3$N and the SnFe$_9$ domains detected in the pristine EC (see Section 3.5). Indeed, no clear peaks ascribed to $FeO_x$ and $SnO_y$ phases are revealed in the WAXD patterns of $FeSn_{0.5}$-$CN_l$ 900/Gr (see Figure 5). As reported in Table 2, the surface of the samples *"Gr support"* and *"$FeSn_{0.5}$-$CN_l$ 900/$Gr_A$"* does not comprise N. The N 1s profiles of both *"$FeSn_{0.5}$-$CN_l$ 900/Gr, Beginning"* and *"$FeSn_{0.5}$-$CN_l$ 900/Gr, End"* reveal four components (see Figure 6), which are assigned in accordance with the literature [65]: (i) pyridinic N @ B.E. (binding energy) ≈ 398.1 eV; (ii) N bound to Fe species @ B.E. ≈ 399.5 eV; (iii) pyrrolic N @ B.E. ≈ 400.5 eV; and (iv) graphitic N @ B.E. ≈ 401.6 eV. The N 1s profile of *"$FeSn_{0.5}$-$CN_l$ 900/Gr, End"* reveals one more N component, ascribed to N-$O_x$ species and centered at *ca.* 402.7 eV [65]. As for the N 1s profiles, the ageing under the X-ray beam of the XPS instrumentation gives rise to the following phenomena: (i) the overall intensity of the N 1s multiplet increases by *ca.* 20%; (ii) the intensity of both the pyridinic N and of the N bound to Fe species does not change; (iii) the intensity of the pyrrolic N and of the graphitic N is increased; and (iv) a new peak appears, ascribed to N-$O_x$ species. The high-res O 1s profiles reveal up to four components (see Figure 6), which are ascribed to [54,66]: (i) oxides of Fe and Sn @ B.E. ≈ 530.4 eV; (ii) C=O species @ B.E. ≈ 531.1 eV; (iii) C-OH or C-O-C species @ B.E. ≈ 532.3 eV; and (iv) chemisorbed $O_2$ or $H_2O$ @ B.E. ≈ 535.7 eV. The ageing under the X-ray beam of the O 1s profile results in: (i) a slight increase in the intensity of the peak ascribed to oxides, that becomes better resolved; and (ii) a slight decrease in the intensity of the peak ascribed to C-O-C / C-OH species.

This evidence can be rationalized as follows. It is expected that at the beginning of the XPS measurements some loosely-bound carbon, possibly bound to oxygen atoms in C-OH and C-O-C groups, partially covers both some of the N species and the oxide species that cover the $Fe_3N$ and $SnFe_9$ nanoparticles (see Section 3.5 and Figure 5). The energy provided by the X-ray beam of the XPS instrumentation likely triggers the following events:

- The loosely-bound carbon is oxidized by the oxygen atoms provided by the oxide species; the latter act as a catalyst for this decomposition process; as a result, $CO_2$ or other volatile species are produced and are then removed from the surface of the EC.
- The removal of the loosely-bound C raises the surface content of the N species found immediately underneath, that are identified as the pyrrolic N and the graphitic N (see Table S2). On these bases we can conclude that: (i) some N species (*i.e.*, those that are not affected by the ageing process, namely pyridinic N and N bound to Fe) are more easily exposed on the EC surface; accordingly, they are expected to play a more important role in the ORR; and (ii) some other species are less easily exposed on the EC surface (*i.e.*, pyrrolic N and graphitic N) and probably provide a lower contribution to the ORR.
- Some fraction of N is also oxidized by the oxygen atoms provided by the oxide species; however, since N is well-stabilized in the system (either in the CN *"shell"*, and especially in the *"coordination nests"*, or onto the $Fe_3N$ nanoparticles) it gives rise to $N-O_x$ species and is not removed from the EC surface.

It is also important to highlight that since the $N_x$-Fe component is not affected by the energy provided by the X-ray beam, we can assume that the N-Fe interactions (that are most likely localized at the interface between the $SnFe_9$ or the $Fe_3N$ nanoparticles and the *"coordination nests"* of the CN *"shell"*) are strong. This evidence suggests that *"coordination nests"* are able to stabilize effectively Fe species, that are expected to play an important role in the ORR mechanism (see Section 3.7, below). The Gr support does not include metals (see Table 1); furthermore, **A** etches completely Fe and Sn from the surface of the activated EC (see Table 2). Consequently, the high-res O 1s profiles of both the Gr support and of $FeSn_{0.5}$-$CN_l$ 900/$Gr_A$ do not exhibit any oxygen component ascribed to oxides. Finally, the relative contributions of the other components of the O 1s peak (*i.e.*, carbonyl groups, hydroxyl/ether groups, and chemisorbed oxygen/water) are very similar for all the samples. This is a further proof that the proposed synthetic route gives rise to a CN *"shell"* exhibiting a high degree of graphitization, where the chemical states of oxygen are very similar to those revealed by the Gr support.

*3.7 CV-TF-RRDE studies*

The proposed ECs are studied by CV-TF-RRDE both in acid (0.1 M $HClO_4$) and in alkaline (0.1 M KOH) media. The resulting profiles are shown in Figure 7.

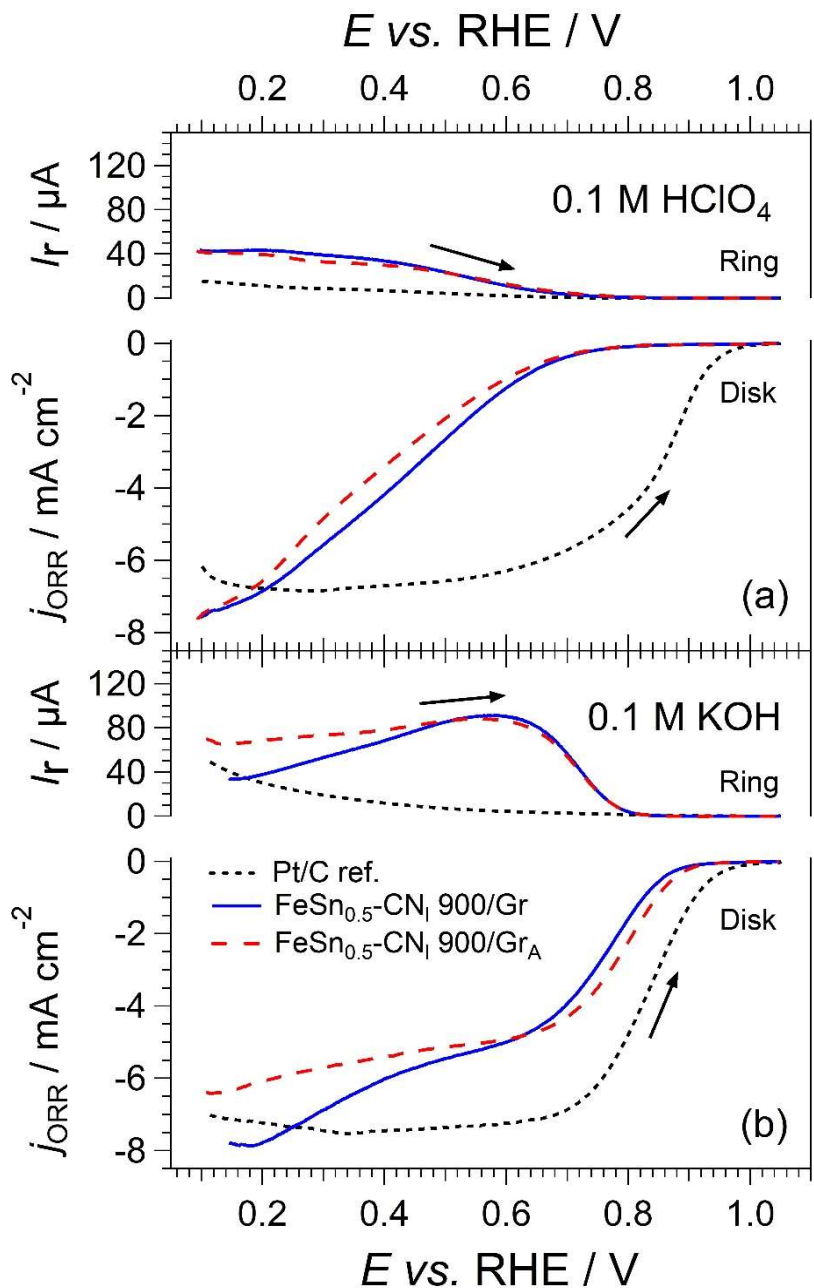

**Figure 7.** CV-TF-RRDE traces under a pure oxygen atmosphere of the Gr-supported ECs. Cell filled with: 0.1 M $HClO_4$ (a); or 0.1 M KOH (b). T = 298 K, sweep rate = 20 mV s$^{-1}$, electrode rotation rate 1600 rpm, $P_{O_2}$ = 1 atm. The ohmic drops and the capacitive currents are corrected as reported elsewhere [11,67].

Equation (1) is adopted to evaluate n, *i.e.*, the average number of electrons that are exchanged during the ORR:

$$n = \frac{4 I_{ORR}}{I_{ORR} + \frac{I_R}{N}} \qquad (1)$$

where: (i) $I_R$ is the ring current; (ii) $I_{ORR}$ is the current ascribed to the ORR process, as measured on the RRDE disk; this latter figure is calculated multiplying the $j_{ORR}$ values displayed on Figure 7 (*i.e.*, the geometric current density in the ORR detected on the RRDE disk) by the geometric area of the RRDE disk, $A_{Disk} \approx 0.196$ cm$^2$; and (iii) N = 0.39 is the collection efficiency of the RRDE ring; its value is determined experimentally as described elsewhere [68]. The profiles of n *vs.* E are shown in Figure 8(a1) and Figure 8(b1). The kinetic current densities in the ORR, $j_k$, are determined by correcting $j_{ORR}$ to account for the diffusion phenomena [27,68]. The $j_k$ values are then adopted to prepare the Tafel plots of the ORR, that are displayed in Figure 8(a2) and Figure 8(b2).

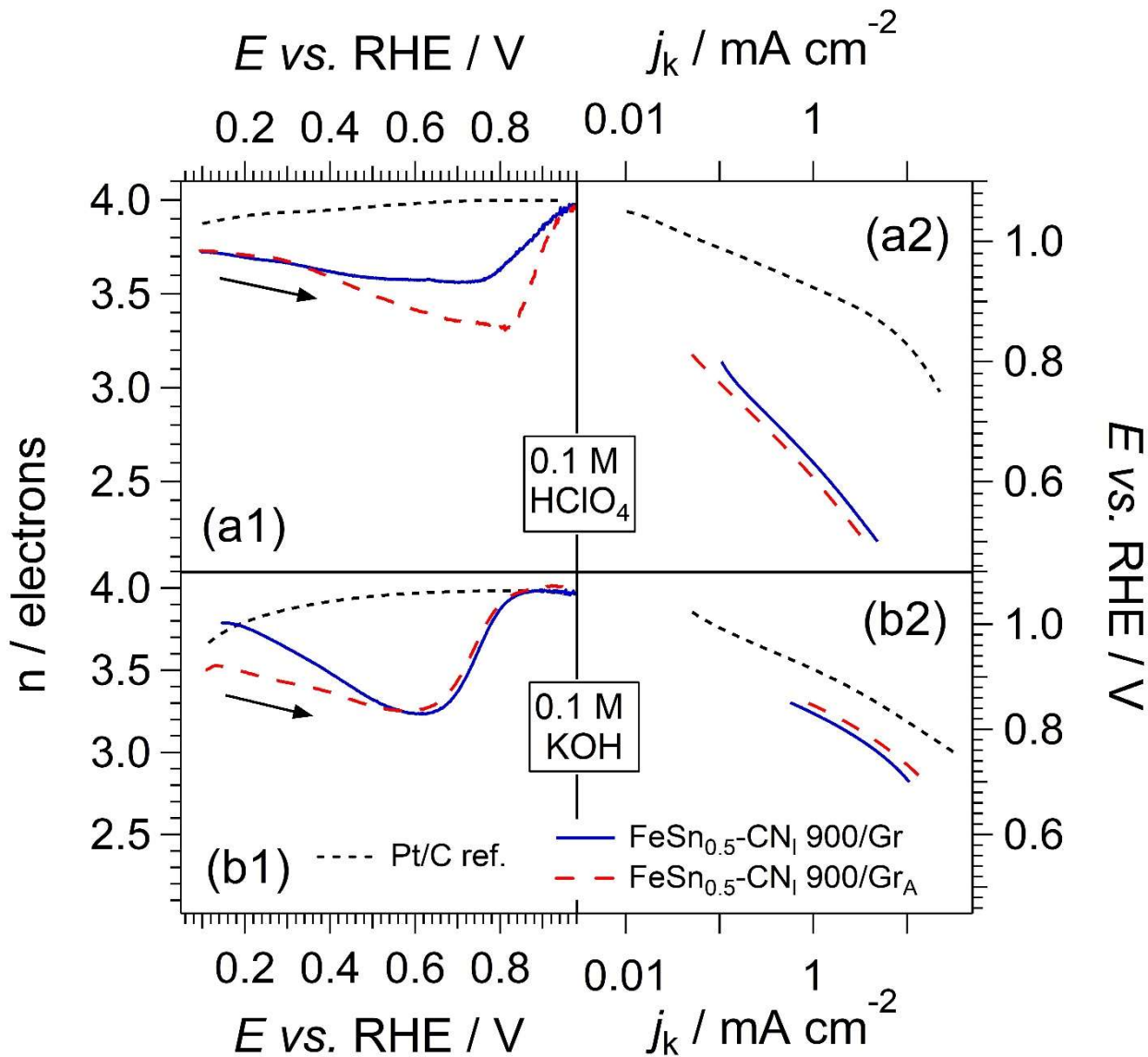

**Figure 8.** Number of electrons exchanged during the ORR (n) in a pure oxygen atmosphere. Cell filled with: 0.1 M $HClO_4$ (a1) and 0.1 M KOH (b1). Tafel plots of data displayed in Figure 7. Cell filled with: 0.1 M $HClO_4$ (a2) and 0.1 M KOH (b2). The experimental conditions are reported in the caption of Figure 7.

In this work, Figure 9 summarizes the figures of merit that are adopted to gauge the EC performance in the ORR. Such figures of merit are: (i) the number of electrons exchanged during the ORR at E = 0.3 V *vs.* RHE (n\*); and (ii) the onset potential, E($j_{5\%}$), which is obtained as described in the caption of Figure 9 [27].

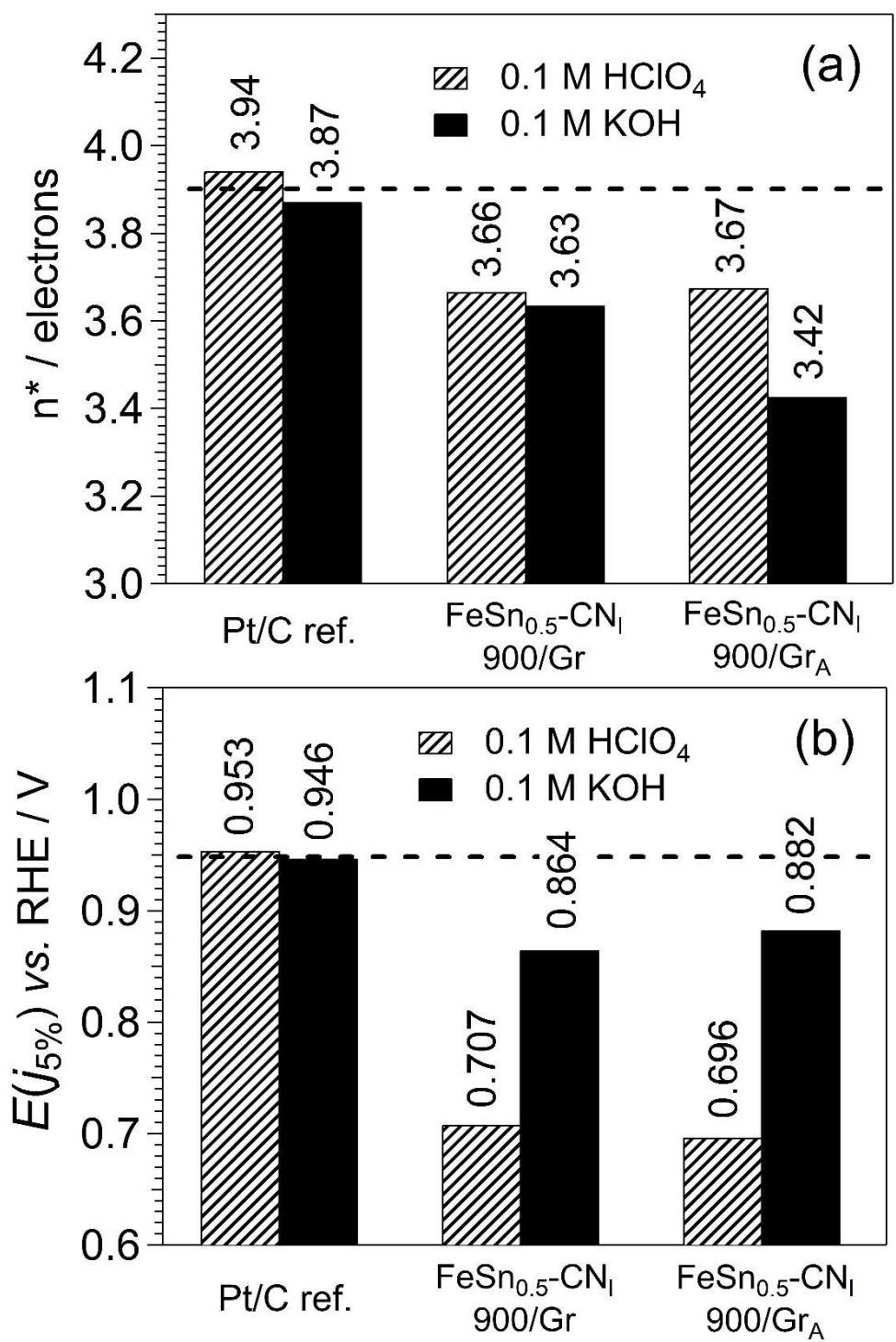

**Figure 9.** Figures of merit describing the performance of the Gr-supported ECs. n* – number of electrons exchanged during the ORR at E = 0.3 V *vs.* RHE (a); onset potential, E($j_{5\%}$) (b); E($j_{5\%}$) is the electrode potential corresponding in the ECs to a $j_{ORR}$ equal to 5% of the maximum ORR limiting current density measured for the Pt/C ref. at *ca.* 0.3 V *vs.* RHE in the same conditions.

The results reported in Figure 7, Figure 8 and Figure 9 are interpreted taking into consideration the framework developed to correlate the physicochemical properties and the electrochemical performance in the ORR of other *"Pt-free"* ECs where the active sites are embedded into a CN *"shell"* matrix covering a conductive *"core"* [27]. The following trends are detected:

- Figure 7 reveals that the ORR overpotentials, $\eta_{ORR}$, increase in the following order: Pt/C (pH = 1; $\eta_{ORR}$ ~ 275 mV) ≤ Pt/C (pH = 13; $\eta_{ORR}$ ~ 285 mV) < ECs (pH = 13; $\eta_{ORR}$ ~ 350 mV) << ECs (pH = 1; $\eta_{ORR}$ ~ 530 mV). In the proposed ECs, the impact of the pH value on $\eta_{ORR}$ is much larger than in the case of the Pt/C ref.
- **A** has an opposite impact on $\eta_{ORR}$ depending on the pH value. Indeed, at pH =1, **A** raises $\eta_{ORR}$ by a few mV; on the other hand, at pH = 13 **A** reduces $\eta_{ORR}$ by more than 15 mV (see also the values of E($j_{5\%}$) reported in Figure 9).
- At pH = 1, disk currents do not reveal any clear diffusion-limited plateau (see Figure 7(a)); on the other hand, at pH = 13 a diffusion-limited plateau is detected (see Figure 7(b)). The latter becomes *"flatter"* upon **A**; correspondingly, $I_R$ values are slightly increased.
- With respect to the Pt/C ref., the Tafel slopes of the proposed ECs: (i) at pH = 13, they are very similar; and (ii) at pH = 1, they are significantly larger, on the order of 180 mV·decade$^{-1}$ or more.
- n* decreases in the following order: Pt/C ref > pristine EC ≈ activated EC. n* is not strongly affected by the pH value, and slightly decreases upon **A**.

The above evidence is rationalized admitting that the ORR mechanism of the proposed ECs is very different with respect to that of the Pt/C ref. The ORR is bottlenecked by the first transfer of electrons from the EC to the dioxygen [10]. At pH = 1, such an electron transfer can only occur through a conventional *"inner-shell"* process, where dioxygen must undergo direct adsorption on the active sites [12,13]. In the case of the Pt active sites of the Pt/C ref., such a process is very facile and yields the observed low values of $\eta_{ORR}$ ($\eta_{ORR} \approx 280$ mV). In the case of the proposed *"Pt-free"* ECs, such a process is likely very difficult, thus yielding the observed high values of $\eta_{ORR}$ ($\eta_{ORR} \approx 530$ mV). If we consider that at pH = 1, in the proposed *"Pt-free"* ECs: (i) the observed Tafel slopes are much larger than the *"ideal"* value of 120 mV·decade$^{-1}$ [69]; and (ii) no diffusion-limited plateau is detected, it is hypothesized that this first rate-determining step is strongly hindered by the mass transport of reactants and products to/from the active sites. At pH = 13, the processes that enable the first ORR electron transfer in the Pt/C ref. are very different in comparison with those occurring in the proposed *"Pt-free"* ECs [12]. In the case of the Pt/C ref., the first electron transfer still occurs through a conventional *"inner-shell"* process, and it is observed that the pH value does not affect significantly $\eta_{ORR}$ ($\eta_{ORR} \approx 280$ mV both at pH = 1 and at pH = 13). In the case of the proposed *"Pt-free"* ECs, the first electron transfer likely occurs mainly through an *"outer-shell"* process engaging a dioxygen molecule centered on a cage of water molecules [12,27]. The latter process is expected to involve most of the EC surface, and not very specific active sites; accordingly, it is much more facile in comparison with the *"inner-shell"* process occurring at pH = 1. These assumptions are justified if we consider that for the proposed ECs, at pH = 13: (i) the values of the Tafel slope closely match those of the Pt/C ref.; (ii) a diffusion-limited plateau is detected at E < 0.7 V *vs.* RHE, similarly to the Pt/C ref. (see Figure 7); finally, in the proposed ECs $\eta_{ORR}$ is significantly reduced, from *ca.* 530 mV (pH = 1) to *ca.* 350 mV (pH = 13). On these bases, we can deduce that: (i) the mass transport of reactants and products no longer hinders the ORR kinetics; and (ii) instead, at the lowest $\eta_{ORR}$ the ORR is modulated by the $O_2$ adsorption isotherm. Indeed, in the proposed ECs the Tafel slope increases progressively as $\eta_{ORR}$ is raised from *ca.* 60 mV·decade$^{-1}$ (corresponding to a Temkin isotherm) to *ca.* 120 mV·decade$^{-1}$ (corresponding to a Langmuir isotherm) [11,70].

The ORR performance in the alkaline medium can be correlated with the physicochemical properties of *"Pt-free"* ECs by considering the figure of merit indicated as *"Specific Surface Activity"*, shortened as *"SSA"*[27]. SSA is defined as the ORR kinetic current at E = 0.85 V *vs.* RHE ($i_{k,ORR@0.85V}$) divided by the area of the active sites of the EC [27]. The area of the active sites of the Pt/C ref. is the surface area of Pt nanoparticles supported thereon, as the latter are largely more efficient to promote the ORR in comparison with the carbon support. The surface area of the Pt nanoparticles is measured by CO stripping experiments [71]. A precise determination of the area of the active sites of the proposed ECs is very difficult since there are not any universally accepted *"probes"* that allow for a selective detection of ORR active sites based on Fe. This shortcoming is mitigated in the proposed *"Pt-free"* ECs, as in the alkaline medium most of the surface is expected to be involved in the ORR (see above). Accordingly, SSA is defined as the ratio between $i_{k,ORR@0.85V}$ and the total surface of the EC as determined by nitrogen physisorption techniques (see Figure 3). The values of SSA are reported in Table 4.

**Table 4.** SSA of the *"Pt-free"* ECs in the alkaline environment

| *Electrocatalyst* | *SSA / µA·cm$^{-2}$* |
|---|---|
| **FeSn$_{0.5}$-CN$_l$ 900/Gr** | 0.504 |
| **FeSn$_{0.5}$-CN$_l$ 900/Gr$_A$** | 0.661 |
| **CoSn$_{0.5}$-CN$_l$ 900/Gr$_A$** | 0.194 [72] |
| **NiSn$_{0.5}$-CN$_l$ 900/Gr$_A$** | 0.023 [72] |
| **Pt/C ref.** | 473[a] |

[a] Value referring to the Pt nanocrystals supported on the Pt/C ref.; additional information reported in [27].

The SSA value of the Pt/C ref. is almost three orders of magnitude larger or more in comparison with the proposed *"Pt-free"* ECs. This is compatible with a vastly easier adsorption of O$_2$ on Pt-based active sites (see above). The ECs labeled *"CoSn$_{0.5}$-CN$_l$ 900/Gr$_A$"* and *"NiSn$_{0.5}$-CN$_l$ 900/Gr$_A$"* are obtained exactly as the EC *"FeSn$_{0.5}$-CN$_l$ 900/Gr$_A$"*; the only difference is that in the former two ECs the *"active metal"* is Co and Ni, respectively. As a first approximation, the other physicochemical features of these three *"Pt-free"* ECs are very similar. Additional details on *"CoSn$_{0.5}$-CN$_l$ 900/Gr$_A$"* and *"NiSn$_{0.5}$-CN$_l$ 900/Gr$_A$"* are reported elsewhere [72]. The SSA of *"Pt-free"* ECs is mostly modulated by the *"active metal"*. In particular, SSA is raised as the *"active metal"* is changed in the order Ni → Co → Fe, following the increase in oxophilicity of the metals [55,73,74]. The latter feature promotes the first approach of oxygen to the EC surface, that is crucial in modulating the first, rate-determining electron transfer process [70]. The available data is not sufficient to identify clearly the exact contribution of the Sn *"co-catalyst"* in the ORR mechanism of the *"Pt-free"* ECs reported in this work. Nevertheless Sn (that is found on the surface of FeSn$_{0.5}$-CN$_l$ 900/Gr$_A$ as SnO$_x$ species, as shown in Figure S5) is assumed to increase the EC hydrophilicity. If this is the case, the first *"outer-shell"* electron transfer from the EC surface to a dioxygen molecule centered on a cage of water molecules [12] would be facilitated, lowering $\eta_{ORR}$. It is highlighted that SSA is raised as the dark features of the pristine EC revealed in Figure 4 are etched upon **A**. Accordingly, it is deduced that the latter dark features, that correspond to SnFe$_9$ and Fe$_3$N domains (see Section 3.5), do not play a crucial role in the ORR. On these bases, it is reasonable to admit that in the proposed ECs the ORR active sites are mostly located on Fe- and Sn- species stabilized in *"coordination nests"* based on C and N. In the pristine EC, such active sites are mainly found at the interface between the CN *"shell"* matrix and the SnFe$_9$ and Fe$_3$N domains. As the latter are etched upon **A**, the active sites become more easily accessible and free from inert contaminants, resulting in the observed increase in SSA (see Table 4). Finally, since the HR-TEM study of the activated EC does not reveal any dark domain (see Figure 4) it can be envisaged that the Fe- and Sn- species making up the active sites are very small, with a size lower than 1-2 nm. In the acid environment the picture outlined above is likely slightly changed: upon **A** a slight increase in $\eta_{ORR}$ is revealed, that corresponds to a modest lowering of E($j_{5\%}$) (see Figure 9). This result is interpreted admitting that in the acid environment the SnFe$_9$ and Fe$_3$N domains play a more pronounced role in the ORR mechanism, slightly promoting the first *"inner-shell"* electron transfer to dioxygen.

In the proposed ECs n is on the order of *ca.* 3.2-3.6, while in the Pt/C ref n is higher, almost 4; furthermore, while in the proposed ECs n decreases as E is raised, the reverse trend is detected in the Pt/C ref (see Figure 8). These findings are interpreted considering that, on Pt active sites, the ORR occurs almost exclusively with a *"direct"* 4-electron mechanism that yields $H_2O$. At low potentials (E < 0.4 V *vs.* RHE) the Pt active sites become progressively more clogged with adsorbates, that inhibit the simultaneous adsorption of the two atoms of an $O_2$ molecule on adjacent Pt sites. This latter is a prerequisite for the direct reduction of $O_2$ to $H_2O$ [10,11]. Instead, a fraction of the incoming $O_2$ molecules only succeeds to undergo adsorption on one Pt site; in this case, $H_2O_2$ is yielded upon the exchange of only 2 electrons [10,11]. In the case of the proposed ECs, the ORR is expected to take place in accordance with a very different, two-step mechanism. (i) In the first step, two electrons are exchanged and $O_2$ undergoes reduction yielding either $H_2O_2$ (pH = 1) [10] or $HO_2^-$ (pH = 13) [75]. (ii) In the second step either intermediate goes through a second two-electron reduction process, that gives rise to $H_2O$ [10]. (iii) A single-step dissociative adsorption of $O_2$ with the exchange of 4 electrons cannot be excluded. (ii) and (iii) are increasingly facilitated as the overpotentials associated to the corresponding phenomena are raised (namely, as E is reduced).

Figure 7(b) reveals that in the pristine EC, at pH = 13, as E is lowered below *ca.* 0.6 V *vs.* RHE: (i) the disk current is progressively increased *"below"* the level of the diffusion-limited plateau; and (ii) correspondingly, $I_R$ is lowered. This evidence is easily interpreted in the framework described above. (i) The *"real"* $j_{ORR}$ value associated to the diffusion-limited plateau is achieved at E ≈ 0.6 V *vs.* RHE, and mostly corresponds to the first, two-electron step of the ORR; the $HO_2^-$ intermediate is then expelled from the electrode layer and later detected on the RRDE ring, giving rise to the maximum value of $I_R$. (ii) as E is lowered below 0.6 V *vs.* RHE, a progressively increasing fraction of the $HO_2^-$ intermediate is further reduced to $H_2O$ and does not manage to escape the porous EC spread on the RRDE disk. (iii) These additional reduction events give rise to the exchange of additional electrons, that are detected on the RRDE disk and raise $j_{ORR}$. Upon **A,** the total *"external"* specific area of the EC is increased (see Figure 3 and Figure 4). Accordingly, the $HO_2^-$ intermediates are better capable to escape the electrode layer on the RRDE disk and are more easily detected on the RRDE ring, raising $I_R$ and lowering both n and n* (see Figure 8(b1) and Figure 9(a)). In the proposed ECs, the trends detected for both n and n* at pH = 1 upon **A** are quite similar to those highlighted at pH = 13 (see Figure 8(a1) and Figure 9(a)). This evidence further suggests that in both the proposed ECs the ORR takes place with a (2 x 2) electron mechanism.

**4. Conclusions**

This report discloses the successful application of the proposed synthetic route to obtain *"Pt-free"* ECs exhibiting a *"core-shell"* morphology and comprising a hierarchical graphene-based support *"core"*. In addition, it is elucidated the complex interplay between the physicochemical features of the proposed *"Pt-free"* ECs and the performance in the ORR.

The proposed ECs include graphene platelets as the *"core"* support, covered with a CN *"shell"* that is rough, microporous and cratered. The latter *"shell"* incorporates the active sites, that consist of Fe- and Sn- species stabilized in *"coordination nests"* of C- and N-based ligands. The *"active metal"* and the *"co-catalyst"* adopted in the present report are Fe and Sn, respectively [41]. $FeSn_{0.5}\text{-}CN_l$ 900/Gr, the *"pristine EC"*, goes through an *"activation process"* **A** that yields the $FeSn_{0.5}\text{-}CN_l$ 900/Gr$_A$ (the *"activated EC"*). **A** modulates significantly the physicochemical properties of the material. In summary, **A**:

- Further promotes the graphitization of the CN *"shell"*; most heteroatoms are etched, raising the $n_H/n_C$, $n_O/n_C$ and $n_N/n_C$ by *ca.* 4-8 times; this is expected to minimize the ohmic losses associated to the transport of electrons from the active sites to the external circuit.

- Etches a significant fraction of Sn and Fe; this is further witnessed by the disappearance in the activated EC of the $SnFe_9$ and $Fe_3N$ domains revealed in the pristine EC by HRTEM, ED, and WAXD. However, after **A**, it is likely that these surviving Sn and Fe species are better stabilized in C- and N-based ligands of the *"coordination nests"*.
- Raises the external specific area of the ECs, improving the accessibility of the active sites upon the removal of mostly ORR-inert $SnFe_9$ and $Fe_3N$ domains.

In the pristine EC it is shown that ageing under the X-ray beam of the XPS instrumentation: (i) does not affect the pyridinic N and the N bound to Fe species; and (ii) etches the loosely-bound carbon covering the pyrrolic N and the graphitic N. This allows to hypothesize that: (i) pyridinc N and N-Fe species are better exposed on the EC surface, playing a more important role in the ORR; and (ii) the Fe-N interactions are strong, further suggesting that N ligands are a crucial component of the *"coordination nests"* stabilizing the Fe species in the ECs. In the proposed ECs the ORR active sites: (i) likely consist of the Sn and Fe species stabilized in the *"coordination nests"* of the CN *"shell"* based on C- and N- ligands; and (ii) are not located on the $SnFe_9$ and $Fe_3N$ domains; on the contrary, as the latter domains are etched, the specific surface activity (SSA) in the ORR is raised (see Section 3.7).

It is also found that the adoption of Fe as the *"active metal"* gives rise to a promising performance in the ORR; this is interpreted considering the high oxophilicity of Fe. Indeed, the latter facilitates the approach of $O_2$ to the EC surface, that triggers the rate-determining first electron transfer. In the proposed ECs, the ORR takes place mostly with a (2 x 2) electron mechanism. At pH = 1, the first rate-determining electron transfer is mediated by a conventional *"inner-shell"* process. Instead, at pH = 13 the first electron transfer is associated with an *"outer-shell"* phenomenon; the latter is much more facile, and results in much lower $\eta_{ORR}$ values (as pH:1→13, $\eta_{ORR}$: 530 mV → 350 mV). In this study, the highest ORR performance is revealed by FeSn$_{0.5}$-CN$_l$ 900/Gr$_A$; at pH = 13, its E($j_{5\%}$) is equal to 0.882 V, less than 65 mV lower in comparison with the Pt/C ref. in the same conditions.

In conclusion, this work indicates that hierarchical Gr supports can be used in the preparation of ORR ECs: (i) exhibiting a *"core-shell"* morphology; and (ii) including active sites stabilized in *"coordination nests"* of the CN *"shell"*. This report: (i) paves the way for the synthesis of new, even more performing ECs based on graphene, that take advantage as much as possible of the outstanding properties of this two-dimensional material; and (ii) improves our understanding of this family of materials, in the continuing efforts to devise highly active *"Pt-free"* ECs for the ORR.

The *"Instruments and methods"*, and the experimental details on the *"Electrochemical experiments"* are described in detail in the Electronic Supplementary Information (ESI).

**Acknowledgments**

The research leading to these results has received funding from: (a) the European Commission through the Graphene Flagship – Core 1 project [Grant number GA-696656]; and (b) the Strategic Project of the University of Padova *"From Materials for Membrane-Electrode Assemblies to Electric Energy Conversion and Storage Devices – MAESTRA"* [protocol STPD11XNRY]. V.D.N. thanks the University Carlos III of Madrid for the *"Càtedras de Excelencia UC3M-Santander"* (Chair of Excellence UC3M-Santander). The authors also acknowledge Massimo Colombo for useful discussions.